\documentstyle[11pt,temp]{article}
\def\mum{\,\mu{\rm m}}

\begin{document}
\noindent
{\LARGE {\bf Unveiling Dust-enshrouded Star Formation in the Early Universe: 
a Sub-mm Survey of the Hubble Deep Field}}

\vspace*{0.2in}

\noindent
David Hughes$^{\ast}$, 
Steve Serjeant$^{\ddagger}$, 
James Dunlop$^{\ast}$, 
Michael Rowan-Robinson$^{\ddagger}$, 
Andrew Blain$^{\S}$, 
Robert G. Mann$^{\ddagger}$, 
Rob Ivison$^{\ast}$, 
John Peacock$^{\ast}$, 
Andreas Efstathiou$^{\ddagger}$, 
Walter Gear$^{\dagger}$, 
Seb Oliver$^{\ddagger}$, 
Andy Lawrence$^{\ast}$, 
Malcolm Longair$^{\S}$, 
Pippa Goldschmidt$^{\ddagger}$,
Tim Jenness$^{\star}$.

\vspace*{0.1in}
\noindent
{\sf $\ast$ Institute for Astronomy, University of Edinburgh, Royal
Observatory, Edinburgh EH9 3HJ, UK\\
$\ddagger$ Astrophysics Group, Imperial College, Blackett Lab.,
Prince Consort Rd., London SW7 2BZ, UK \\
$\S$ Cavendish Astrophysics Group, Cavendish Laboratory, Madingley Road, Cambridge CB3 0HE, UK \\
$^{\dagger}$ Mullard Space Science Lab., University College London,
Holmbury St.\,Mary, Surrey RH5 6NT, UK\\
$^{\star}$ Joint Astronomy Centre, 660 N. A'ohoku Place, Hilo, Hawaii
96720, USA 
}

\vspace*{-0.0in}
\noindent
\rule{6.5in}{0.5mm}

\vspace*{0.05in}
\noindent 
{\bf The advent of sensitive sub-mm array cameras now allows a proper
census of dust-enshrouded massive star-formation in very distant
galaxies, previously hidden activity to which even the faintest optical
images are insensitive. We present the deepest sub-mm
survey of the sky to date, 
taken with the SCUBA camera on the James Clerk Maxwell
Telescope and centred on the Hubble Deep Field. The high 
source density found in this 
image implies that the survey is confusion-limited 
below a flux density of 2~mJy. However, within the central 80 arcsec radius
independent analyses yield 5 reproducible sources with ${\bf S_{850\mum} >
2}$~mJy which simulations indicate can be ascribed to individual galaxies. 
We give positions and flux densities for these, and furthermore
show using multi-frequency photometric data that the brightest sources
in our map lie at redshifts ${\bf z \simeq 3}$. 
These results lead to integral 
source counts which are completely inconsistent with a no-evolution model, 
and imply that massive star-formation activity continues at redshifts ${\bf >
2}$. The combined brightness of the 5 most secure sources in our map is 
sufficient to account for 30 -- 50\% of the previously unresolved sub-mm
background, and we estimate statistically that the entire background
is resolved at about the 0.3~mJy level.
Finally we discuss possible optical identifications and redshift 
estimates for the brightest
sources. One source appears to be associated with an extreme starburst
galaxy at $\bf z \simeq 1$, whilst the remaining four appear to lie 
in the redshift range $\bf 2 \leq z < 4$. This
implies a star-formation density over this redshift range that is at
least five
times higher than that inferred from the ultraviolet output of HDF galaxies.}
 
 \noindent
 \section*{Understanding Star-Formation at High Redshift}
 
 Recent years have seen the first meaningful attempts to determine the
 global star-formation history of the Universe, using the combined leverage
 provided by deep redshift surveys ({\it e.g.} the Canada France Redshift 
 Survey$^{\bf 1}$) reaching $z \simeq 1$, and the 
 statistics of Lyman-limit galaxies$^{\bf 2}$ 
 at higher redshifts in, for example, the Hubble Deep Field
(HDF)$^{\bf 3,4,5}$.
 The results$^{\bf 6}$ imply that the star-formation and 
 metal-production rates were about 10 times greater at $z \simeq 1$ than
 in the local Universe, that they peaked at a redshift in the range $z 
 \simeq 1 \rightarrow 1.5$  and that they declined to values comparable to 
 those observed at the present day at $z \simeq 4$.

 This conclusion, derived from optical-UV data, may however be misleading,
 because the absorbing effects of dust within distant galaxies
 undergoing massive star-formation may have distorted our
 picture of the evolution of the high-redshift Universe in two ways.
 First, the star-formation rate (SFR) in {\it known}  high-redshift objects is 
 inevitably under-estimated unless some
 correction
 for dust obscuration is included in deriving the rest-frame UV
 luminosity.
 Second, it is possible that an entire population of heavily dust-enshrouded
 high-redshift objects, as expected in some models of elliptical galaxy
 formation$^{\bf 7}$, have gone undetected in the optical/UV surveys. 
 The extent of the former remains controversial$^{\bf 8,9,10,11}$
 while the possibility of the latter has 
 until now been impossible to investigate.
 
 At high redshifts ($z > 1$), the strongly-peaked far-infrared (FIR)
 radiation
 emitted by star-formation regions in distant galaxies is redshifted into
 the sub-mm waveband, and the steep spectral-index of this emission longward of
 the peak, at $\lambda \simeq 100\mum$ in the rest-frame, results in  
 a large  negative K--correction which is sufficient at sub-mm wavelengths
 to offset the dimming of galaxies due to their cosmological distances. 
Consequently the flux density of a galaxy at $\lambda \simeq 800\mum$ with 
fixed
 intrinsic FIR luminosity is expected to be roughly constant at all
 redshifts in the range $1 \leq z \leq 10^{\bf 12,13,14}$.
 
 This ease of access to the young Universe has already been exploited 
 through successful pointed sub-mm observations of known high-redshift sources 
 including lensed objects (IRAS\,F10214+4724$^{\bf 15}$
 and the Cloverleaf quasar$^{\bf 16}$),
 radio galaxies$^{\bf 14,17,18}$
 and quasars$^{\bf 19,20}$. These studies 
 have demonstrated the potential of sub-mm cosmology
 and have shown that in at least some high-redshift galaxies, 
 dust-enshrouded star-formation is proceeding at a rate of $\gg 100 
 \,{\rm M_{\odot} yr^{-1}}$, substantially greater than the more modest
 star-formation rates ({\it e.g.} on average $\sim 1 - 5 \,h^{-2} {\rm M_{\odot} yr^{-1}}$) 
displayed by Lyman-break galaxies$^{\bf 3}$. 

 With the recent commissioning of the sensitive sub-mm array camera SCUBA on the
 JCMT$^{\bf 21}$ it is now possible to conduct unbiased sub-mm selected
 surveys$^{\bf 22}$ and quantify the amount of star-formation  activity in
 the young universe by observing directly the rest-frame 
 FIR emission from dust in high-redshift galaxies.
In this paper we describe the first results from an ultra-deep sub-mm survey
centred on the HDF.
 
 \section*{A Sub-mm Survey of Hidden Starformation in the Hubble Deep Field}

 Recent ISOCAM observations of the HDF 
 at 6.7$\mum$ and 15$\mum$ have confirmed that the strong evolution seen 
 in the IRAS galaxy population at low redshifts$^{\bf 23,24}$ 
 continues out to redshifts of order unity$^{\bf 25,26}$. 
 Such mid-infrared
 studies can, however, provide no constraints at higher redshift.
 In contrast an 850$\mum$ survey is predicted 
 to be completely dominated by sources at $z \ge 1$, and the number of 
 detectable sources is very sensitive to the high-redshift evolution
 of the dusty starburst population. In particular, a SCUBA survey of the HDF
 complete to a flux density limit $S_{850\mum}>2$~mJy would be 
 expected to detect $< 0.1$ galaxies if there is no cosmic evolution, 
 $< 1$ galaxy if the evolution mirrored the Madau
 curve$^{\bf 4}$, but at least 2 sources if the number density and luminosity 
 of infrared starburst galaxies continued to evolve strongly out to 
 $z \simeq 2$, and substantially more sources with $z>2$ if the 
 population continued to evolve or stayed constant at higher
redshifts$^{\bf 7,12,27,28,29}$.
 
 We chose to centre this deep 850$\mum$ survey  
 on the HDF, not only because the SCUBA field of view of 
 $\simeq 6$ arcmin$^2$ is
 well matched to the area of the HDF, but also to maximise the possibility of finding
 optical/IR/radio counterparts and redshifts for any sub-mm 
 sources which are detected. Currently there exist over $20$ 
 spectroscopic redshifts for galaxies at $z>2$
 within the HDF, while the availability of deep photometric data$^{\bf 30}$ in the
 $U_{300}$, $B_{450}$, $V_{606}$ and $I_{814}$ bands facilitates the
 estimation of photometric redshifts for other galaxies in the field.

 Simultaneous diffraction-limited images of the HDF at 850$\mum$ and
 450$\mum$ were taken with SCUBA$^{\bf 21}$ on the 15-m
 James Clerk Maxwell Telescope.  A total of 50 hours integration
 between January\,5th and February\,13th 1998 
 were centred at $12^{\rm h}36^{\rm m}51^{\smash{\rm s}}\cdot 20$ 
 $+62^\circ 12' 52\smash{{}''}\cdot 5$ (J2000) with occasional
 offsets 25\,arcsec south, east and west to aid the discrimination of
 real and spurious sources. The sub-mm data were taken under exceptional
 atmospheric conditions, with a median $850\mum$ sky opacity
 $\tau_{\rm 850\mum} = 0.16$. 
 Sky subtraction was performed using on-array chopping in Right
 Ascension in order to minimise the chop throw (important for accurate
 sky subtraction), to maximise the reclaimable signal-to-noise ratio for
 detected sources, and to minimise the number of negative off-beams
 arising from unknown sources well outside the primary field of view.
 We experimented with chopping in azimuth, but, at least at the declination
 of the HDF, this yielded no significant noise improvement over chopping
 in RA. Finally, to ensure that no significant source would be missed due
 to an unfortunate coincidence with the off-beam of another brighter source, 
 the length of chop throw was varied, approximately half
 the observations (29 hr) adopting an RA chop-throw of 30 arcsec, and the
 remainder (21 hr) an RA chop-throw of 45 arcsec. As discussed 
 below, this approach proved invaluable
 both for source confirmation, and for the separation of real and confused 
 sources.  The 850$\mu$m data, with an angular resolution 
of 14.7 arcsec FWHM,  covers an area of approximately
9\,arcmin$^2$ and, due to the variation in the density of bolometer
samples across the map, has a noise at the 
periphery approximately double its value at the map centre. 
The 850$\mu$m image in Figure\,1  shows a circular field, within a
radius of 100 arcsec from the map centre, and  
reaches a 1$\sigma$ noise level of 0.45~mJy/beam. This image
represents by far the deepest sub-mm map ever taken.  

 \section*{Sub-mm source extraction and confusion}
 
 Because the noise increases with radius from the map centre, 
 sources were only sought within the central 80 arcsec radius of
 the image. The map shown in Figure\,1  displays 58 distinct
 peaks, the majority of which are noise. For Gaussian
 filtered white noise, 1\% of the peaks exceed $3.3\sigma$ in
 amplitude$^{\bf 31}$, and so a flux density of 1.5~mJy (at the map
centre) is the practical
 detection threshold for real sources; the map contains 7 such objects.
 The use of two different chops and the effect of telescope nodding is 
 to produce a convolving beam with four negative sidelobes. 
The signature of a source
 is therefore very different from noise, and this fact can be used to
 identify real sources and to deconvolve the map down to some flux
 density limit.
 Deconvolution also allows the flux in the sidelobes of a source to be
 reclaimed, thereby enhancing the signal-to-noise ratio of the detected
 sources.
In order to investigate whether these peaks correspond to single, or
 blended sources, simulations of 
 random source distributions with plausible number counts have been
 carried out. The 14.7-arcsec beam is sufficiently
 broad that an ideal noise-free map would in fact never
 contain more than about 20 peaks within the 5.6\,arcmin$^{2}$ map,
 independent of the true density of sources. Alternatively,
 the observed source density is about one source per 12
 beam areas; both arguments indicate that source confusion
 must become important at the limit of our 850$\mum$ map.
 
It is thus possible that at least some of
the apparent sources in the map could consist of
emission from more than one object, and this
is a particular concern for the weaker sources
with $S_{850\mum}\simeq 2$~mJy. One way of isolating
such cases is optical identifications, as discussed
below; if there is only a single candidate identification, the source
cannot be a blend, since each member of the blend would have 
a separate optical counterpart.

Another approach is to note that confusion is only a
serious problem when there is a blend of one or
more sources of similar flux, and that in such
cases the apparent source will usually be
significantly broader than the telescope beam.
This breadth means that the apparent source position
will be less stable under the addition of noise
than if the source is dominated by a single unresolved object.
We have therefore taken the conservative approach
of identifying sources in the full data-set that appear in both 
the 30-arcsec and 45-arcsec chop images, and only keeping those
whose positions agree to better than 3 arcsec.
Tests on simulated source
fields with realistic number counts show that
this procedure should succeed in giving a clean
sample of the sources brighter than 2~mJy in the central 80 arcsec
radius of the image, each dominated by a single object.
 The positions of these 5 sources are given in table 1, 
 together with their 850-$\mu$m flux densities.

The simultaneous 450$\mum$ image covers 75\%
 of the useful area mapped at 850$\mum$ and,  
 despite the excellent observing conditions, which resulted in
 a $\rm 1\sigma$ rms noise signal of 7~mJy/beam at 450$\mum$,
 no significant detections were obtained. 
 
 \section*{Number Counts and the Sub-mm Background}
 
 The map shown in Figure\,1 can be used to determine the form of the number
 counts at 850$\mum$ fainter than the limit
 of 2~mJy at which individual sources can be selected with some
 confidence. Fainter sources combine to raise the rms fluctuations in the
 map beyond what is expected purely from noise.
 There is a long
 tradition in radio and X-ray astronomy of extracting
 faint counts from such information using `$P(D)$' analyses$^{\bf 32}$, 
 although the present dataset is unusual in that both random noise and
 confusion noise are of similar amplitude. The approach adopted here is to
 focus upon the
 distribution of signal-to-noise ratios for the peaks of the map in
 Figure\,1 (i.e. the distribution of fluxes for all apparent `sources').
 By generating synthetic maps with different number
 counts, it is possible to estimate what range of true counts is
 consistent with the observed distribution.
 
 We have not explored the full parameter space, but some examples are 
 illustrated in Figure\,2.
 Empirically, it is clear that there is an
 excess of peaks in the range 0.8 to 1.5~mJy, and
 this requires a substantial density of sources
 at about this flux-density level. The observed peak flux-density
distribution is
 matched reasonably well by a source density of about
 $7000\,\rm deg^{-2}$ brighter than 1~mJy,
 which corresponds to the observed density of
 brighter sources, extrapolated with a Euclidean
 count slope, with the major caveat that this
 number assumes an unclustered source distribution.
 If in fact the faint sub-mm sources are high-redshift
 starbursts, it is not implausible that they are strongly clustered 
 on scales of several arcsec$^{\bf 33,34}$.
 For a given surface density of sources, this increases
 the background fluctuations and so the above figure
 should probably be treated as an upper limit.
 
 The counts must continue to flux densities somewhat fainter
 than 1~mJy, but the present data do not have the sensitivity to 
 estimate where the inevitable break from the Euclidean slope occurs.
 This is best constrained by asking at what flux density the
 extrapolated count exceeds the background.  By summing the
 flux densities in Table 1, a lower limit to the background
 contributed by discrete sources of 20~mJy/5.6 arcmin$^2$ is found,
 equivalent to
 to $\nu I_{\nu} = 1.5 \times 10^{-10}\,\rm Wm^{-2}sr^{-1}$, or 
 approximately half the background estimate reported by Puget 
{\it et al.}$^{\bf 35}$.
 There is, however, evidence in our data, specifically by continuing the
 deconvolution until the residual noise is statistically
 symmetric, or using the cumulative counts to 1~mJy derived above, 
 that the true background
 contributed by discrete sources may be up to a factor of two higher than
 this, essentially identical to the original estimate of Puget et al.,
 and consistent with more than 50\% of the revised background estimates 
 at $850\mum^{\bf 36, 37}$ 
 which suggest $\rm \nu I_{\nu} = 5.0 \pm 4 \times 10^{-10} 
 \, \rm W m^{-2} sr^{-1}$. 
 The faint counts must therefore flatten by a
 flux density of about 0.3~mJy, otherwise even this
 background estimate would be exceeded.

 \section*{Photometric Observations, Spectra and Redshift Estimation}
 
 Additional photometric observations during February 1998 
 at the centroid position of
 HDF850.1 confirmed the detection of the brightest
 sub-mm source
 with detections at 1350$\mum$ of $2.1 \pm 0.5$~mJy and at
 850$\mum$ of $7.0 \pm 0.4$~mJy. These data, together with a
 450$\mum$ 3$\sigma$ upper limit of 21~mJy/beam provide
 a robust photometric estimate of the redshift of the source.
 In Figure\,3 the expected flux density ratios at sub-mm and
 mm-wavelengths are plotted as a function of redshift 
 for a range of models, typical of dusty, starforming galaxies,
 which are consistent with the observed optical to sub-mm spectra of
 Arp220, one the most heavily
 enshrouded local starburst galaxies$^{\bf 38,39}$, and the 
 high-$z$ starburst/AGN
 IRAS F10214+4724. 
 The relevance of these models to galaxies in the high-$z$
 universe is reinforced by noting that the measured sub-mm and
 mm-wavelength flux density ratios of high-$z$ AGN$^{\bf 14,40}$ 
 lie close to or within the bounds of the models.
 
 The photometric redshift for HDF850.1, determined from the  
 1350/850$\mum$ flux density ratio, lies within the range $ 2.5 < z < 9$.
 This strong constraint is supported by its non-detection at 450$\mum$
 which provides an upper limit to the 450/850$\mum$ flux density ratio 
 and hence a lower limit to its redshift of $z>3$.
 Less stringent, but similar high redshift limits can be estimated for all
 sub-mm sources detected in the HDF by arguing that their
 non-detection at 15$\mum$ at a 3$\sigma$ level of $\sim20\mu$Jy
 (Oliver {\em et al.} -- in preparation)
 implies a lower limit of $z \sim 2$ 
for sources at 850$\mu m$ brighter than 2~mJy,
 assuming a starburst galaxy model$^{\bf 38}$, or $z > 1.5$ assuming
the observed SED of the extreme starburst galaxy Arp\,220.
 The radio-FIR correlation$^{\bf 41}$ 
 yields lower limits of $z=1.75$ and $z=2.75$ respectively for
 2~mJy and 7~mJy sources detected at 850$\mum$, but not detected at
 8.5\,GHz $^{\bf 42}$ at a 5$\sigma$ flux limit of 9$\,\mu$Jy.
 The above data demonstrate that deep sub-mm surveys provide an 
 efficient means of 
 identifying  a population of star-forming galaxies at redshifts $>2$.
 
 \section*{Optical associations with sub-mm sources in the HDF}
 
 Given the rest-frame optical-FIR ratios typical of luminous starburst
 galaxies, the high-redshift SCUBA-selected 
 galaxies are not necessarily expected to be present in the optical HDF, 
 despite its depth. Nevertheless, we briefly discuss in turn plausible 
 associations for the 5 most secure sub-mm sources in the HDF (see
Figure\,4), estimating photometric
redshifts$^{\bf 43}$, $z_{\rm ph}$, for those galaxies  without spectroscopic 
redshifts extended to include limits
where galaxies are not detected in all four HDF bands. 

Our approach is as follows. For each SCUBA source, we have considered as
a potential optical counterpart all galaxies detected in the HDF 
whose distance from the SCUBA source lies within the 90\%
confidence limit of the sub-millimetre source position listed in Table 1.
For each candidate we have then calculated 
the probability that a galaxy with 
such an optical magnitude (or brighter)
could lie so close to the SCUBA position by chance, and also
the probability that a galaxy with the observed redshift (or higher)
could lie so close to the SCUBA position by chance.
Note that these probabilities are often substantially higher than the raw
Poisson probabilities$^{\bf 44}$. 
This is due to the combined effect of 
the rather large uncertainty in the SCUBA positions, and the high
surface density of galaxies at the limit of the optical HDF image, which
together essentially guarantee that (with the exception of HDF850.1) 
every SCUBA source will have
at least one optical indentification candidate at the limit of the HDF image.
Finally we have investigated whether any of the apparently most probable optical
identifications can in fact be clearly rejected on the basis of the SED 
constraints discussed above.
 
 {\bf HDF850.1}\ As shown in Fig.\,4, this source lies 1.0 arcsec
 from galaxy 3-577.0 in the optical HDF catalogue$^{\bf 30}$
 which has a tentative spectroscopic redshift of $z=3.36$ (ref. 45),
 and which has been claimed$^{\bf 46}$ to be
 gravitationally lensed by the foreground $I_{814}(AB)=24$ 
 elliptical galaxy 3-586.0 which lies at $1.0 \leq z
 \leq 1.2$ (refs. 43,47,48).
 More recently, a 3.5$\sigma$  detection (6.3$\mu$Jy) at 8.5-GHz source has been 
 associated with 3-586.0 (ref. 42). 
 Based on its magnitude the probability that 3-577.0 is a chance association
 with HDF850.1 is 0.33, while based on its redshift (which we estimate is
 $z_{\rm ph}=3.1$) the probability (calculated
 from the surface density of $z_{\rm
 ph}>3$ galaxies in photometric
 redshift catalogues$^{\bf 43,47,48}$) is only 0.20.
 For 3-586.0 the probabilities are in fact comparable (0.29 and 0.49
 respectively), but the non-detection of HDF850.1
 at 15$\mum$ ($S(3\sigma) < 23\mu$Jy) is strongly inconsistent (by almost
 two orders of magnitude) with the observed SEDs of any known galaxy
 (including Arp\,220) if placed at 
 the `low' redshift of 3-586.0. Moreover, as discussed above, the mm/sub-mm flux
 ratios also indicate that $z > 2.5$, and
 Figure\,5 shows
 that the observed spectrum of HDF850.1
 agrees well with that expected for a starburst galaxy at redshift
 $z \sim 3$.
 We note that the random probability of being 2 arcsec
 from one of the radio sources$^{\bf 42}$ is only 0.03. If the
 radio source really is associated with 3-586.0, and HDF850.1 
 with 3-577.0, then these seemingly incompatible probabilities are best
 explained by assuming that  3-577.0 is indeed being gravitationally
 lensed by 3-586.0, thereby amplifying its rest-frame FIR flux, and
 increasing its chances of being detected at 850$\mum$ by
 SCUBA. The amplification would, however, need to be fairly substantial to
 explain the statistics, implying a massive lens;
 3-586.0 may be the only visible member of a fainter group of galaxies.
 
 {\bf HDF850.2}\  lies just beyond the edge of the HDF making an assessment of
 possible optical associations difficult since the $I_{814}$ band 
 Hubble Flanking Field (HFF) image only reached a depth of $\sim25$
 mag. HDF850.2 is 4.3 arcsec from the $z_{\rm ph}=3.8$
 galaxy 3-962.0 on the edge of the HDF, but 
 based on its magnitude the probability that 3-962.0 is a chance association
 with HDF850.2 is 0.63, while based on its redshift the probability is
 0.46. As can be seen in Figure 4, there does appear to be a more
 convincing, but also very faint candidate identification within the HFF,
 but at present we possess little useful colour information for this
 object. Therefore, while the non-detection at 15$\mu$m implies a flux density ratio 
 $\rm S(850\mu m/15\mu m) > 190$ consistent with the SED of
 a starburst galaxy at $z > 2$, we are unable to make an unambiguous
 optical association.
 
 {\bf HDF850.3}\ lies only 1.3 arcsec from 1-34.2 which is an asymmetric
 galaxy with $I_{814}(AB)=24.5$ for which we estimate $z_{\rm ph}\sim 1.95$.
 This is a moderately convincing identification since 
 based on its magnitude the probability that 1-34.2 is a chance association
 with HDF850.3 is only 0.29 (although based on its redshift the probability 
 is 0.52) and its estimated redshift is consistent with 
 a non-detection or marginal detection at 15$\mum$. The next nearest object
 is 1-34.0, a $I_{814}(AB)=21$ galaxy at a distance of only 1.5 arcsec. 
 For this galaxy the random probabilities are 0.12 and 0.60 respectively
 but with a tentative spectroscopic redshift of 0.49, and photometric
 redshift estimates in the range $0.26 \leq z_{\rm ph} \leq 0.68$, this object
 can be confidently rejected as a possible identification given its
 non-detection at 15 $\mu m$, 450$\mu m$ and radio wavelengths. We note
 also that 1-34.0 shows no obvious signs of starburst activity at optical
 wavelengths, and appears to be a relatively undisturbed spiral galaxy.
 Should the identification with 1-34.2 prove to be erroneous we note for
 completeness that two $z_{\rm ph}\sim 3.9$ galaxies, the nearer being 1-27.0,
 and the further 1-31.0,  lie within 4 arcsec of HDF850.3. Based on 
 magnitude the probability that these are chance associations is 0.63,
 while based on redshift it is 0.3.
 
 {\bf HDF850.4}\ lies less than an arcsec from 2-339.0, an $I_{814}(AB)=23$
 galaxy for which photometric redshifts have been determined in the range
 0.74-0.88 (refs. 43,47,48). This is a convincing 
 identification since the based on its magnitude the probability that
 2-339.0 is a chance association with HDF850.4 is only 0.07 (based on 
 its redshift the probability is 0.44).
 Moreover, this is the one case for which 
 the 850$\mu$m source can be plausibly associated 
 with an ISOCAM detection at 15$\mu$m, which yields a flux density ratio 
 $\rm S(850/15\mu m) \sim 16$. This is in fact exactly the value expected 
 from the observed SED of Arp220 if placed at $z \simeq 1$. This supports an 
 identification with 2-339.0, and emphasises the usefulness of 
 constraining the redshift 
 using  the $\rm S(850/15\mu m)$ ratio. Furthermore we 
 note that this optical galaxy is clearly
 disturbed, as would be expected for an extreme starburst galaxy, 
 providing further circumstantial evidence that it is the
 correct optical identification.
 However, should this identification prove erroneous we note that there are 3
 galaxies (2-294.0, 2-315.0, 2-319.0) 
 with $z_{\rm ph}>3$ within 3.5 arcsec of HDF850.4.

{\bf HDF850.5} is located in a sparsely-populated region of the HDF, but
is only 0.9 arcsec away from the $I_{814}(AB)\sim29$ galaxy 2-426.0, for
which we estimate a photometric redshift of $z_{\rm ph}=3.2$. 
This is a moderately convincing identification;
based on its magnitude the probability that it is a chance coincidence
is 0.46, but based on its redshift it is a more impressive 0.16.
This high-redshift
association is consistent with the lack of a 15$\mu$m detection at that
position, but we note the difficulty of estimating photometric redshifts
for such faint HDF galaxies.
Finally we note that 7 other faint ($I_{814}(AB) > 27.5$) galaxies lie
within 3.5 arcsec of HDF850.5 (see Figure 4), but these objects are
significantly more likely to be chance coincidences, and in any case all
also have photometric redshifts $z_{\rm ph} > 2$.
 
In summary, HDF850.4 appears to be associated with a disturbed starburst galaxy
at $z_{\rm ph} \simeq 1$, HDF850.3 with an asymmetric 
galaxy at $z_{\rm ph} \simeq 2$, and 
HDF850.1 and HDF850.5 have relatively unambiguous
associations with galaxies at $z_{\rm ph} \simeq
3$. For HDF850.2 we find a possible identification within the HDF 
at $z \simeq 4$, and an alternative (also faint) candidate in the HFF
which can be reasonably expected to lie at $z > 2$.
Finally we note that we are clearly 
unable to rule out the possibility that the true optical counterparts
of a few of these sources may be too faint for detection even in the HDF.

 \section*{Dust masses, star-formation rates and star-formation density at
 high redshift}
 
 The photometric redshifts and suggested optical 
 identifications for the sub-mm sources 
 are consistent with the expectation that all galaxies
 detected in the 850$\mu$m
 survey of the HDF down to a flux limit $S_{850\mum} = 2$~mJy
 should have redshifts $z \geq 1$.
 Given this, and the flat flux-density--redshift relation between $z = 1$
 and $z = 10$, the dust-enshrouded SFRs and the dust
 masses for all 5 reliable sources can be estimated, independent of their
 precise redshifts.
 The results are given in table 2, and indicate that these sources
 are extremely dusty and have SFRs, when determined from the submillimetre
 data, that are similar to, or exceeding, that of the local
 ultraluminous starburst galaxy Arp220. Note that the calculated SFRs
 are sensitive to the assumed IMF and stellar mass-range (and can in the
 extreme increase and decrease by a factor of $\sim 3$).
 More striking is the comparison of the FIR SFRs with those calculated
 from the rest-frame UV luminosities. The FIR method gives SFRs on average
a factor $\sim 300$ larger. It has been shown that optical SFRs,
estimated from Balmer emission line luminosities in Lyman-break galaxies 
at $\rm z\sim 3$, are larger than the UV SFRs by factors of 2-15. 
This upward correction, due to attenuation by dust$^{\bf 9,10}$, still
would 
require a further factor of $\sim 50$ to explain the higher FIR SFRs.
A similar situation has been observed in the local universe
where the ratio of FIR and H$\alpha$ luminosities in ultra-luminous IR
starburst galaxies (ULIRGs) is $\sim 60
\times$ larger than that in disk galaxies$^{\bf 49}$, 
suggesting that in young starbursts most OB stars are still deeply buried in
their opaque parent clouds. 
The submillimetre sources we have seen are then quite
typical of local ULIRGs, but their relevance to the
overall cosmic star formation history depends on their space density.

The small uncorrected UV SFRs ($< 1\, h^{-2} \rm M_{\odot}  yr^{-1}$, table 2)
for the optical counterparts of the submillimetre sources are reasonable, 
given that these galaxies are typically 
a few magnitudes fainter in the rest-frame UV, 
possibly due to greater dust obscuration,
than the population of Lyman-break galaxies at $z \sim 3.5$ which have
SFRs of $\sim 2\, h^{-2} \rm M_{\odot} yr^{-1}$.
Alternatively it may be that the less secure identifications are
erroneous, and that some of the sub-mm sources may have true optical 
counterparts below the detection limit on the HST HDF image, and therefore
probably at $z \geq 5$.

 By summing the FIR SFRs and dividing by the appropriate 
 cosmological volume, a first, conservative 
 estimate of the level of dust-enshrouded star-formation
 rate in the high redshift Universe can be made using observations that
 are  insensitive to the obscuring effects of dust.
 For illustrative purposes, it can be reasonably assumed that four of
the five sources (all but HDF850.4)
 lie in the redshift interval $2 < z <
 4$, in which case a lower-limit to the dust-enshrouded star-formation rate density is 
 $0.21\, \rm \,h\, M_{\odot} yr^{-1} Mpc^{-3}$ (assuming $q_0 = 0.5$) 
 at $z \simeq 3$. This datum is plotted
 in figure\,6, where it can be compared with the optically-derived 
 star-formation history of the Universe$^{\bf 3,4}$, the dust-corrected 
 star-formation history predicted from the evolution of radio-loud 
 AGN$^{\bf 50}$ and that inferred from the 
 metal-production rate as determined from the observed column densities 
 and metallicities in QSO absorbers$^{\bf 51,52}$. 
 
 If the redshift distribution extended beyond $z = 4$, the
 mean redshift would increase,  but due to increased cosmological volume 
the star-formation rate density would decrease in
 such a way that the data would remain consistent with the curves
 shown in Figure\,6. For example, assuming a redshift range $2 < z < 5$ yields 
 a star-formation density of $0.16\, \,h\, \rm M_{\odot} yr^{-1}
 Mpc^{-3}$ at $z \simeq 3.5$. 
 
Finally, we emphasize that the SCUBA datum plotted in Figure 6 
is in fact rather robust. For example, should 
our proposed identification for HDF850.3 (1-34.2) prove to have a spectroscopic redshift significantly
lower than $z = 2$, the impact on Figure 6 will be to lower the
$z \simeq 3$ SCUBA datum by only 20\%. 
However, the impact of even one of these sources lying at still higher
redshift is rather dramatic; if one of the
2-3 mJy sub-mm sources we have detected actually lies at $z > 4$, this
would yield a
star-formation density of $0.1\, \,h\, \rm M_{\odot} yr^{-1}
 Mpc^{-3}$ in the redshift range $4 < z < 6$, thus keeping the star-formation
density essentially constant out to $z \simeq 5$.

In summary, this deep  submillimetre survey of the HDF demonstrates 
that a significant fraction ($>80\%$) of the star-formation activity in the
high-redshift universe may have been missed in previous optical
studies. Four of the five 
 brightest submillimetre sources alone provide a density of dust-enshrouded
 star-formation at $z>2$ which is at least a factor of $\simeq 5$
 greater than that deduced from Lyman limit systems$^{\bf 3}$. 
 The extent to which even this is an under-estimate depends on the number
 of sources fainter than $S_{850} = 2$mJy at comparable redshift.

This unique submillimetre survey of unprecedented sensitivity has
identified a population of high-redshift dusty starburst
galaxies which contribute a significant fraction of the 
extragalactic background at 850$\mum$. 
These observations, together with complementary wider, shallower
submillimetre surveys, are now beginning to provide the first 
true measurement of the starformation history for the early universe, 
unhindered by the attenuating effects of dust.
 The challenge for the future is to follow up these observations, in
particular those of the sub-mJy sources
 which at present can only be detected statistically. 

It is now possible
that some of these objects lie at $z>5$, but demonstrating
 this will require the individual detection of these sources with
 sub-arcsec position errors.  This will require both improved
 noise performance and higher spatial resolution in order to evade the
 confusion limit. Facilities such as the forthcoming generation of 
 sub-millimetre arrays are ideally matched to these key programmes for
 astrophysical cosmology.
 
\newpage

 \section*{References}
 
 \begin{enumerate}
 
 \item{
Lilly, S.J., Le F\'{e}vre, O., Hammer, F., Crampton, D. The
Canada-France Redshift Survey: The Luminosity Density and Star
Formation  History of the Universe to $z\sim1$. {\em Astrophys. J.\/}
{\bf 460}, L1-L4 (1996).
}

\item{
Steidel, C.C., Hamilton, D. Deep imaging of high redshift QSO fields
below the Lyman limit: I. The field of Q0000-263 and galaxies at
z=3.4. {\em Astron. J.\/} {\bf 104}, 941-949 (1992).
}

\item{
Madau, P. {\em et al.\/} High-redshift galaxies in the Hubble Deep
Field: colour selection and star formation history to  $z\sim4$.
{\em Mon. Not. R. Astron. Soc.\/} {\bf 283}, 1388-1404 (1996).
}

\item{
Madau, P. The Evolution of Luminous Matter in the Universe
in `The Hubble Deep Field', eds. Livio,M. {\it et al.}, STScI
Symposium Series, in press (astro-ph/9612157).
}

\item{Connolly, A.J., Szalay, A.S., Dickinson, M. SubbaRao, M.V, 
Brunner, R.J.  The evolution of the global starformation 
history as measured from the Hubble Deep Field.
{\em Astrophys. J. \/}, {\bf 486}, L11-L14 (1997). 
}

\item{
Fall, S.M., Charlot, S., Pei, Y.C. Cosmic emissivity and background
intensity from damped Lyman-alpha galaxies. {\em Astrophys. J. \/}
{\bf 464} L43-L46 (1996).
}

\item{Franceschini, A., Mazzei, P., De Zotti, G., Danese, L., 
Luminosity evolution and dust effects in distant galaxies:
implications for the observability  of the early evolutionary phases.
{\em Astrophys. J.}, {\bf 427}, 140-154 (1994).
}

\item{
Pettini, M., Smith, L.J., King, D.L., Hunstead, R.W., The metallicity
of high-redshift galaxies: the abundance of zinc in 34 damped
Ly-$\alpha$ systems from z=0.7 to 3.4. {\em Astrophys. J.} {\bf 486}, 
665-680  (1997). 
}

\item{
Pettini, M. {\it et al.}
The Spectra of Star Forming Galaxies at High Redshift
in `The Ultraviolet Universe at Low and High Redshift', ed. W. Waller,
AIP Press (1998).
}

\item{
Heckman, T.M., Robert, C., Leitherer, C., Garnett, D.R., van der
Rydt, F. The ultraviolet spectroscopic properties of local starbursts:
implications at high-redshift. {\em Astrophys. J.\/}, submitted
(astro-ph/9803185).
}

\item{
Meurer, G.R., Heckman, T.M., Lehnert, M.D., Leitherer, C.,  Lowenthal, J.
The panchromatic starburst intensity limit at low and high redshift.
{\em Astron. J.\/} {\bf 114}, 54-68 (1997).
}

\item{
Franceschini, A., De Zotti, G., Toffolatti, L., Mazzei, P., Danese,
L. Galaxy counts and contributions to the background radiation
from 1 micron to 1000 microns. {\em Astron. Astrophys. Suppl. Ser.\/}
{\bf 89}, 285-310 (1991).
}

\item{
Blain, A.W., Longair, M.S. Submillimetre cosmology. {\em Mon. Not. R.
Astron. Soc.\/} {\bf 264}, 509-521 (1993).
}

\item{
Hughes, D.H., Dunlop J.S. \& Rawlings, S. High-redshift radio galaxies
and quasars at submillimetre wavelengths: assessing their evolutionary
status. {\em Mon. Not. R. Astron. Soc.\/} {\bf 289}, 766-782 (1997).
}

\item{
Rowan-Robinson, M. {\em et al.} The ultraviolet-to-radio continuum of
the ultraluminous galaxy IRAS F10214+4724. {\em Mon. Not. R. Astron.
Soc.\/}
{\bf 261}, 513-521 (1993).
}

\item{
Barvainis, R., Antonucci, R., Hurt, T., Coleman, P., Reuter, H.-P.
The broadband spectral energy distributions of the Cloverleaf quasar
and IRAS F10214+4724. {\em Astrophys. J.\/} {\bf 451}, L9-L12 (1995).
}

\item{
Dunlop, J.S., Hughes, D.H., Rawlings, S., Eales, S.A. \& Ward, M.J.
Detection of a large mass of dust in a radio galaxy at redshift
$z=3.8$. {\em Nature} {\bf 370}, 347-349 (1994).
}

\item{
Ivison, R.J., {\em et al.} Dust, gas, and evolutionary status of the
radio galaxy 8C 1435+635 at $z=4.25$. {\em Astrophys. J.\/} {\bf 494},
211-217 (1998).
}

\item{
Isaak, K.G., McMahon, R.G., Hills, R.E. \& Withington, S. Observations
of high redshift objects at submillimetre wavelengths.
 {\em Mon. Not. R. Astron. Soc.\/} {\bf 269}, L28-L31 (1994).
}

\item{
Omont, A. {\it et al.}, Continuum millimetre observations of
high-redshift radio-quiet QSOs - II. Five new detections at ${\rm z} > 4$.  
{\em Astron. Astrophys.\/} {\bf 315}, 1-10   (1996). 
}

\item{
Holland W.S. {\it et al.} in 
Advanced Technology MMW, Radio, and TeraHertz Telescopes
Phillips,T.G.,(ed), Proc. of SPIE Vol. 3357, in press (1998)
}

\item{
Smail, I. Ivison, R.J. \& Blain, A.W. A deep sub-millimeter survey of
lensing clusters: a new window on galaxy formation and evolution.
{\em Astrophys. J.\/} {\bf 490}, L5-L8 (1997).
}

\item{
Saunders, W. {\em et al.} The 60-micron and far-infrared luminosity
functions of IRAS galaxies. {\em Mon. Not. R. Astron. Soc.\/}
{\bf 242}, 318-337 (1990).
}

\item{
Lawrence, A. {\em et al.}, The QDOT all-sky IRAS galaxy redshift
survey. {\em Mon. Not. R. Astron. Soc.\/}, submitted. 
}

\item{
Oliver, S.J. {\it et al.}. 
Observations of the Hubble Deep Field
with the Infared Space Observatory - III. Source counts and P(D) analysis.
{\em Mon. Not. R. Astron. Soc.\/} {\bf 289}, 471-481 (1997).
}

\item{
Rowan-Robinson, M. {\it et al.}. 
Observations of the Hubble Deep Field
with the Infared Space Observatory - V. Spectral energy distributions,
starburst models and star formation history.  
{\em Mon. Not. R. Astron. Soc.\/} {\bf 289}, 490-496 (1997).
}

\item{
Blain, A.W., Longair, M.S. Observing strategies for blank-field surveys
in the submillimetre waveband. {\em Mon. Not. R. Astron. Soc.\/} {\bf
279}, 847-858 (1996).
}

\item{
Pearson, C., Rowan-Robinson, M. Starburst galaxy contributions to
extragalactic source counts. {\em Mon. Not. R. Astron. Soc.\/} {\bf
283}, 174-192 (1996).
}

\item{
Hughes, D.H., Dunlop, J.S.  Continuum observations of the
high-redshift universe at sub-millimetre wavelengths, in `Highly
Redshifted Radio Lines', eds. C.Carilli {\it et al.} in ASP
Conf. Ser. (1998).
}

\item{
Williams, R.E. {\em et al.} The Hubble Deep Field: observations, data
reduction, and galaxy photometry. {\em Astron. J.\/} {\bf 112},
1335-1389 (1996).
}

\item{
Bond, J.R., Efstathiou, G. The statistics of cosmic background
radiation fluctuations. {\em Mon. Not. R. Astron. Soc.\/}
{\bf  226}, 655-687 (1987).
}

\item{
Condon J.J. Confusion and flux-density error distributions.
{\em Astrophys. J.\/} {\bf 188}, 279-286 (1974).
}

\item{
Steidel C.C. {\it et al.}
A large structure of galaxies at redshift {$\rm z \sim 3$} and its
cosmological implications,
{\em Astrophys. J.\/}, in press (astro-ph/9708125).
}

\item{
Giavalisco M. {\it et al.}
The angular clustering of Lyman-break galaxies at redshift z=3, 
{\em Astrophys. J.\/}, in press (astro-ph/9802318).
}

\item{
Puget, J.L. {\em et al.} Tentative detection of a cosmic far-infrared
background with COBE. {\em Astron. Astrophys.\/} {\bf 308}, L5-L8 (1996).
}

\item{
Fixen,D.J., Dwek, E., Mather, J.C., Bennett, C.L., Shafer, R.A.
The spectrum of the extragalactic FIR background from the COBE IRAS
observations.  astro-ph/9803021
}

\item{
Guiderdoni, B., Bouchet, F.R., Puget, J.L., Lagache, G. \& Hivon, H.
The optically dark side of galaxy formation. {\em Nature\/} {\bf 390},
257-259 (1998).
}

\item{
Rowan-Robinson, M., Efstathiou, A. Multigrain dust cloud models of
starburst and Seyfert galaxies. {\em Mon. Not. R. Astron. Soc.\/}
{\bf  263}, 675-680 (1993).
}

\item{
Dudley, C.C., Wynn-Williams, C.G. The deep silicate absorption
feature in IRAS 08572+3915 and other infrared galaxies. {\em Astrophys.
J.\/} {\bf 488}, 720-729 (1997).
}

\item{
Ivison, R.J. {\em et al.} 
A hyperluminous galaxy at z = 2.8 found in a deep submillimetre survey
{\em Mon. Not. R. Astron. Soc.\/} in press, (astro-ph/9712161).
}

\item{
Sopp, H.M., Alexander, P. A composite plot of far-infrared versus
radio luminosity, and the origin of far-infrared luminosity in quasars.
{\em Mon. Not. R. Astron. Soc.\/} {\bf 251}, 14P-16P (1991).
}

\item{
Richards, E.A., Kellermann, K.I., Fomalont, E.B., Windhorst, R.A.,
Partridge, R.B. Radio emission from galaxies in the Hubble Deep Field.
{\em Astron. J.\/} submitted (astro-ph/9803343).
}

\item{
Mobasher, B., Rowan-Robinson M., Georgakakis, A., Eaton, N.
The nature of the faint galaxies in the Hubble Deep Field.
{\em Mon. Not. R. Astron. Soc.\/} {\bf 282}, L7-14 (1996).
}

\item{ 
Browne, I.W.A., Cohen, A.M. Quasars near bright
galaxies. {\em Mon, Not. R. Astron. Soc.\/}, {\bf 182}, 181-187 (1978).
}

\item{
Zepf, S.E., Moustakas, L.A., Davis, M. Keck spectroscopy of objects
with lens-like morphologies in the Hubble Deep Field. {\em Astrophys. J.\/}
{\bf 474}, L1-L5 (1997).
}

\item{
Hogg, D.W., Blandford, R., Kundic, T., Fassnacht, C.D., Malhotra, S.
A candidate gravitational lens in the Hubble Deep Field.
{\em Astrophys. J.\/} {\bf 467}, L73-L75 (1996).
}

\item{
Sawicki, M.J., Lin, H., Yee, H.K.C. Evolution of the galaxy population
based on photometric redshifts in the Hubble Deep Field.
{\em Astron. J.\/} {\bf 113}, 1-12 (1997).
}

\item{
Wang, Y., Bahcall, N., Turner, E.L. A catalogue of colour-based redshift
estimates for $z\stackrel{<}{_\sim}4$ galaxies in the Hubble Deep Field.
{\em Astron. J.\/} submitted (astro-ph/9804195).
}

\item{
Leech, K.J. {\it et al.} High-luminosity IRAS galaxies -
II. Optical spectroscopy, modelling of starburst regions and
comparison with structure. 
{\em Mon. Not. R. Astron. Soc.\/} {\bf 240}, 349-372 (1989).
}

\item{
Dunlop, J.S.
Cosmic star-formation history, as traced by radio-source evolution, 
in `Observational cosmology with the new radio surveys', Bremer {\it
et al.} (eds.) 157-164, Kluwer (1998). 
}

\item{
Fall S.M., Pei Y.C.
Obscuration of quasars by dust in damped Lyman-alpha systems.
{\em Astrophys. J.\/} {\bf 402}, 479-492 (1993).
}

\item{
Pei Y.C., Fall S.M. 
Cosmic chemical evolution.
{\em Astrophys. J.\/} {\bf 454}, 69-76 (1995).
}

\item{
Sandell, G. Secondary calibrators at submillimetre wavelengths
{\em Mon. Not. R. Astron. Soc.\/} {\bf 271}, 75-80 (1994).
}

\item{
Jenness T., Lightfoot J. F., Reducing SCUBA data at the James Clerk
Maxwell Telescope, in `Astronomical Data Analysis
Software and Systems VII', 
A.S.P. Conf. Ser. Vol.145, eds. Albrecht R. {\it et al.}, 216-219 (1998).
}

\item{
Hook, R.N., Fruchter, A.S, Variable-Pixel Linear Combination,
in `Astronomical Data Analysis Software and Systems VI',
A.S.P. Conf. Ser. Vol.125, eds. Hunt,G., Payne,H.E., 147- (1997).
}

\item{
Gallego, J., Zamorano, J., Aragon-Salamanca, A., Rego, M. The current
starformation rate of the local universe. 
{\em Astrophys. J.\/} {\bf 455}, L1-L4 (1995).
}

\end{enumerate}

\newpage

\suppressfloats

 \begin{table}
 \caption[]{\scriptsize Positions and flux densities for the 5 most reliable 
 sub-mm sources in the HDF with $S_{850}>2$~mJy. The positions of these 
 sources are reproduced from deconvolutions of both the 30-arcsec and
 45-arcsec chopped images to within 3 arcsec. Positions given for
 each source were obtained from the average of the positions obtained
using the independent SURF and IDL reductions.
The quoted r.m.s. positional uncertainties were derived from the formula
 $\sigma_{\rm pos}=\theta_{\rm beam}/(2S/N)$, where $\theta_{\rm beam}$
 is the FWHM of the beam and $S/N$ is the signal-to-noise
 ratio of the source. A further 0.5~arcsec uncertainty was
 added in quadrature, to account for the standard error in
 absolute pointing derived from measured pointing offsets throughout the
 observations. Flux densities quoted are the average from 3
 independent methods, all of which agreed to within the formal uncertainty.
 Absolute calibration is uncertain to 10\%.}
 \begin{center}
 \begin{tabular}{cccc}
 IAU name   & RA (J2000) & Dec (J2000) &
 S$_{850\mum}$ (mJy)  \\  
 HDF850.1  J123652.3+621226 & 12 36 52.32 ($\pm 0.10$) & +62 12 26.3 ($\pm 0.7$) & $7.0 \pm 0.5$ \\ 
 HDF850.2  J123656.7+621204 & 12 36 56.68 ($\pm 0.20$) & +62 12 03.8 ($\pm 1.4$) & $3.8 \pm 0.7$ \\ 
 HDF850.3  J123644.8+621304 & 12 36 44.75 ($\pm 0.21$) & +62 13 03.7 ($\pm 1.5$) & $3.0 \pm 0.6$ \\ 
 HDF850.4  J123650.4+621316 & 12 36 50.37 ($\pm 0.23$) & +62 13 15.9 ($\pm 1.6$) & $2.3 \pm 0.5$ \\ 
 HDF850.5  J123652.0+621319 & 12 36 51.98 ($\pm 0.25$) & +62 13 19.2 ($\pm 1.8$) & $2.1 \pm 0.5$ \\  
 \end{tabular}
 \end{center}
 \end{table}

\vspace{0.5in}
  
\begin{table}
\caption[]{\scriptsize 
Star-formation 
rates and dust masses of 5 most reliable sub-mm sources in HDF with
$S_{850} >2$~mJy.
Column\,2: Photometric redshifts based on the most probable optical
associations. Column\,3: 60$\mum$ luminosity determined from the
starburst model of M82$^{\bf 38}$ scaled to the 
observed 850$\mum$ flux densities. Columns\,4,5:
Star-formation rates calculated from rest-frame UV (2800{\rm \AA})
and FIR (60$\mum$) luminosities$^{\bf 3,26}$. The UV flux densities at 2800{\rm \AA} are
interpolated from the measured $\rm I_{814}(AB)$ and 
$\rm V_{606}(AB)$ magnitudes$^{\bf 30}$. Column\,6: Dust masses,
assuming $\beta=1.5$, $\rm T=50\,K^{\bf 15}$. An Einstein-de Sitter
cosmology is assumed.}
\begin{center}
\begin{tabular}{cccccc}
  &   & $\rm log_{10}\, L_{60\mum}$ & 
\multicolumn{2}{c}{SFR ($h^{-2} \rm M_{\odot}  yr^{-1}$)} & $\rm log_{10}\, M_{dust}$ \\  
 source & $\rm z_{est}$ & $(h^{-2} \rm L_{\odot})$ & UV & FIR &
$(h^{-2} \rm M_{\odot})$ \\
 HDF850.1  & 3.4 & 12.15 & 0.7 & 311 & 7.88 \\
 HDF850.2  & 3.8 & 11.87 & 0.2 & 161 & 7.62 \\
 HDF850.3  & 2.0 & 11.76 & 2.0 & 127 & 7.52 \\
 HDF850.4  & 0.9 & 11.83 & 0.7 & 142 & 7.50 \\
 HDF850.5  & 3.2 & 11.64 & 0.3 &  95 & 7.36 \\
 \end{tabular}
 \end{center}
 \end{table}

\pagebreak[4]

\newpage
\clearpage


\begin{figure}
\begin{center}
\setlength{\unitlength}{1mm}
\caption[]{\scriptsize 
The 850$\mu$m SCUBA image of the HDF. 
The image shows a radius of 100 arcsec from the map centre
($12^{\rm h}36^{\rm m}51\cdot 20^{\smash{\rm s}}$ 
$+62^\circ 12' 52\cdot 5\smash{{}''}$ - J2000) and is orientated with
North upwards and East to the right. 
Primary flux calibration was performed using Uranus, with secondary 
 calibration against a variety of AGB stars and compact HII
 regions$^{\bf 53}$. The absolute calibration uncertainty is $< 10\%$. 
 The data, taken in jiggle-map mode, were reduced in parallel using two wholly independent methods.
 The first reduction
 used the SCUBA User Reduction Facility (SURF v.1.2)$^{\bf 54}$,
whilst the second reduction was performed with a specially-written IDL
pipeline. Both methods incorporate individual bolometer rms noise
weighting in the map reconstruction. An iterative
 temporal deglitching and spatial sky subtraction was performed. The IDL
 maps were reconstructed using a noise-weighted ``drizzling''
 technique$^{\bf 55}$ and were in excellent agreement with the
independent SURF reconstructions.
Individual subsets of the data were also reduced using both techniques, including 
 the production of separate images with 30 arcsec and 45 arcsec chop throws.
  The centre of the map contains a higher density of bolometer samples
 than the periphery. Consequently the noise is a function of
 position and this variation can be deduced exactly from
 the known jiggle pattern, and is approximated closely
 by a quadratic radial variation $\sigma\propto
 1 + (r/90\,\rm arcsec)^2$ in the central regions.
 In the SURF reduction, the noise has 
 the statistical character of white noise
 filtered with a beam of FWHM 6 arcsec, with an
 rms of 0.65~mJy at the map centre.
 Convolution of the map reduces this noise, but
 possible confusion from faint sources means that it is preferable not to
 broaden the point-source response significantly. As a
 compromise, a further convolution with a
 6-arcsec beam was applied, reducing the rms noise signal to 0.45~mJy
at the map centre.
The noise on the final map therefore has the character of white noise
convolved with a beam of FWHM 8.5 arcsec.
A signal-to-noise image is shown, allowing the significance
of faint potential sources to be judged in a uniform manner, 
although it means that there is a tendency for more sources to
be detected in the central regions.
The analysis of sources was restricted to the central 
80 arcsec radius. Because the observing strategy yields a point-source
response with negative sidelobes at 0.25 of the peak,
the map was CLEANED and restored with a 14.7 arcsec FWHM gaussian beam.
}
\end{center}
\end{figure}


\begin{figure}
\begin{center}
\setlength{\unitlength}{1mm}
\caption[]{\scriptsize
The raw integral number counts for the central 80 arcsec
radius of the map in figure\,1
is shown as the jagged solid line on this plot. 
The flux units here are signal-to-noise, but scaled
to the flux units in the map centre.
This uniform-noise representation allows a
clear demonstration of sources in excess of the
expectation for a pure noise field (dashed line)
above about 0.8~mJy. Synthetic maps were made with
the observed noise properties using a random distribution
of sources having Euclidean counts down to a limit of 0.3~mJy
(although the results are insensitive to this cutoff).
The grey band shows the effect of varying the integral
surface density at 1~mJy between 4000 and 10000 degree$^{-2}$.
}
\end{center}
\end{figure}

\begin{figure}
\begin{center}
\setlength{\unitlength}{1mm}
\caption[]{\scriptsize 
An estimation of redshift using measured 
submillimetre flux densities. 
The hatched area shows the range of submillimetre flux density ratios
as a function of redshift which are constrained by two extreme models of
dusty, starforming galaxies$^{\bf 38}$ (Arp220 and IRAS10214+4724)
and are consistent with
observations of high-z galaxies$^{\bf 12,40}$. 
The solid horizontal lines represent the measured flux ratios 
for HDF850.1 
and, in the case of the $\rm S(1350/850\mum)$ ratio,
the horizontal dotted lines represent errors of $\pm 1\sigma$ on the
observed ratio. The solid shading represents the parameter space
satisfied by the photometric data for HDF850.1, including the non-detection at 450$\mum$,   
and illustrates that the redshift for HDF850.1 probably 
lies between $\rm 2.5 < z < 9$. 
}
\end{center}
\end{figure}

\begin{figure}
\begin{center}
\setlength{\unitlength}{1mm}
\caption[]{\scriptsize Optical associations for the brightest five
submillimetre sources in the HDF. The top-left panel indicates 
the approximate location, and
orientation of each of the $10\times10$ arcsec $I_{814}$-band postage stamps
shown in the following 5 panels. In each postage stamp 
the two large circles represent the 90\% and
50\% confidence limits on the sub-mm positional uncertainty, whilst a small 
circle (1\,arcsec in diameter) has been used to mark the location of the most
plausible optical association (or associations) for each SCUBA source (see text).}
\end{center}
\end{figure}

\begin{figure}
\begin{center}
\setlength{\unitlength}{1mm}
\caption[]{\scriptsize
The observed optical--radio spectral energy distribution of HDF850.1.  
The solid circles represent the SCUBA 850 and 1350$\mu$m detections. 
Non-detections at 8.5\,GHz$^{\bf 42}$, 450$\mu$m (this paper), 
15 and 6.7$\mum$ (Oliver {\it et al.} - in prep.) are shown as open diamonds.  
The solid squares indicate the optical fluxes in the $I_{814}$, 
$V_{606}$ and $B_{450}$ HST bands 
of the $\rm z=3.36$ galaxy 3-577.0, which is a plausible association for HDF850.1.
A starburst galaxy model (solid curve)$^{\bf 38}$  has been 
redshifted to $z=3.36$ and normalised to the 850$\mum$ flux density. 
An additional radio non-thermal synchrotron component (where $F_{\nu} \propto
\nu^{-0.8}$) is scaled to the starburst model at 60$\mum$ using the
radio-FIR correlation$^{\bf 41}$.
The 15$\mum$ upper-limit is consistent with the model SED since 
the contribution from a rest-frame $3.3\mum$ PAH feature, averaged
over the ISOCAM LW3 bandpass, is insignificant.
}
\end{center}
\end{figure}

\begin{figure}
\begin{center}
\setlength{\unitlength}{1mm}
\caption[]{\scriptsize 
The global star-formation history of the universe. Traditionally the
mean {\it comoving} rate of formation of stars in the universe,
$d\rho_{\rm stars}/dt$, has been measured from
the total UV luminosity density of galaxies.
At $z<1$, this was measured by the Canada-France
Redshift Survey of Lilly {\it et al.}$^{\bf 1}$, and
at higher redshifts from the optical HDF data$^{\bf 3}$.
The zero-redshift datum was inferred from local emission-line 
galaxies$^{\bf 56}$. The shaded region shows
the prediction (assuming $h=0.65$) due to Pei \& Fall$^{\bf 52}$
who argued using the observed column densities in QSO absorbers, plus
the low metallicities in these systems, that the star-formation
rate must have peaked between $z=1$ and $z=2$. 
The solid line illustrates what would
happen if the star-formation rate tracked the total output
of radio-loud AGN$^{\bf 50}$. Based on the evidence which
indicates that four of the five brightest sub-mm HDF sources
lie in $2 < z < 4$, we infer a rate about 5 times higher than
that obtained by Madau$^{\bf 3}$, but in good agreement with the
external predictions$^{\bf 50,51}$ of the rate at these epochs.
}
\end{center}
\end{figure}

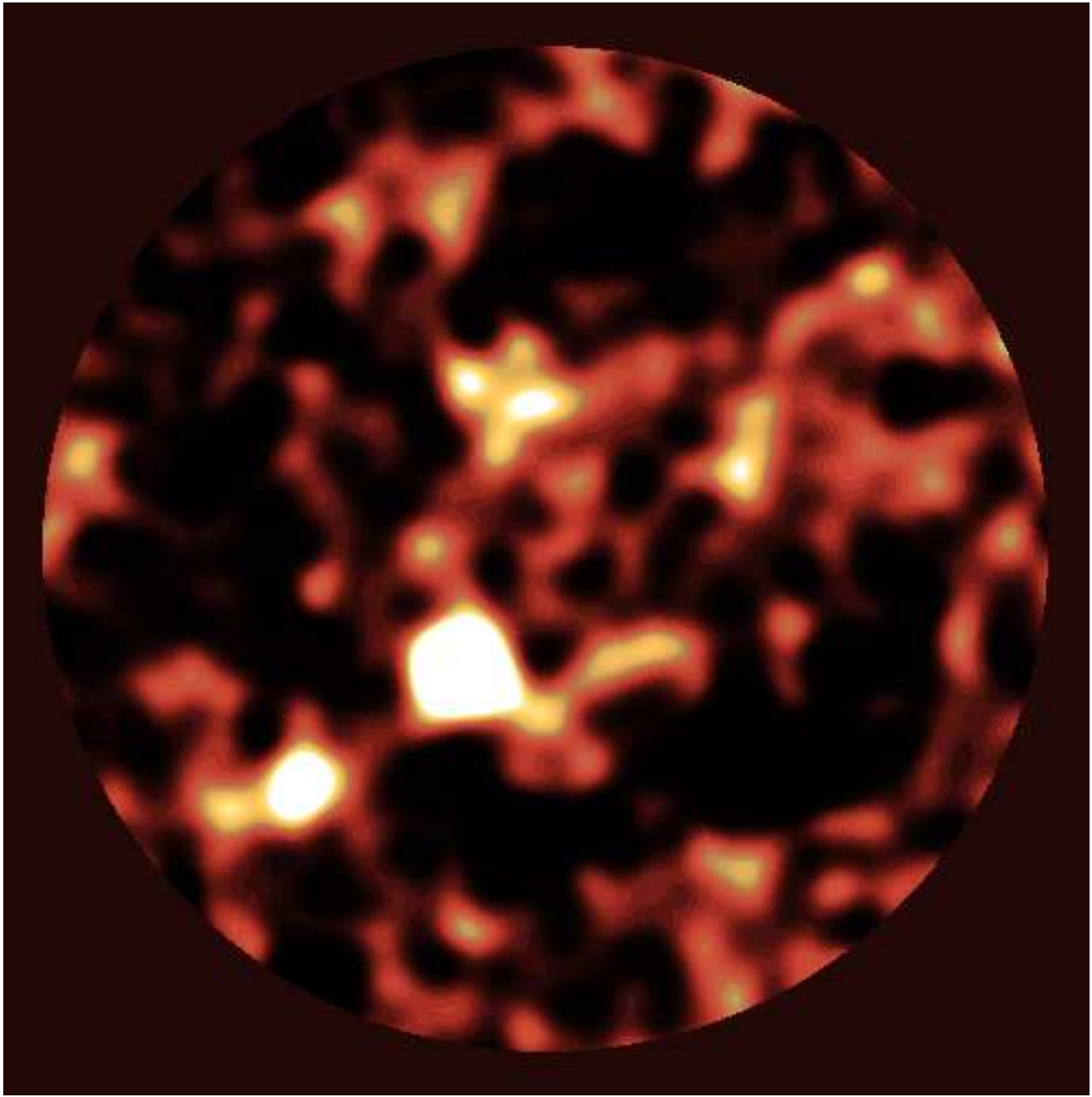
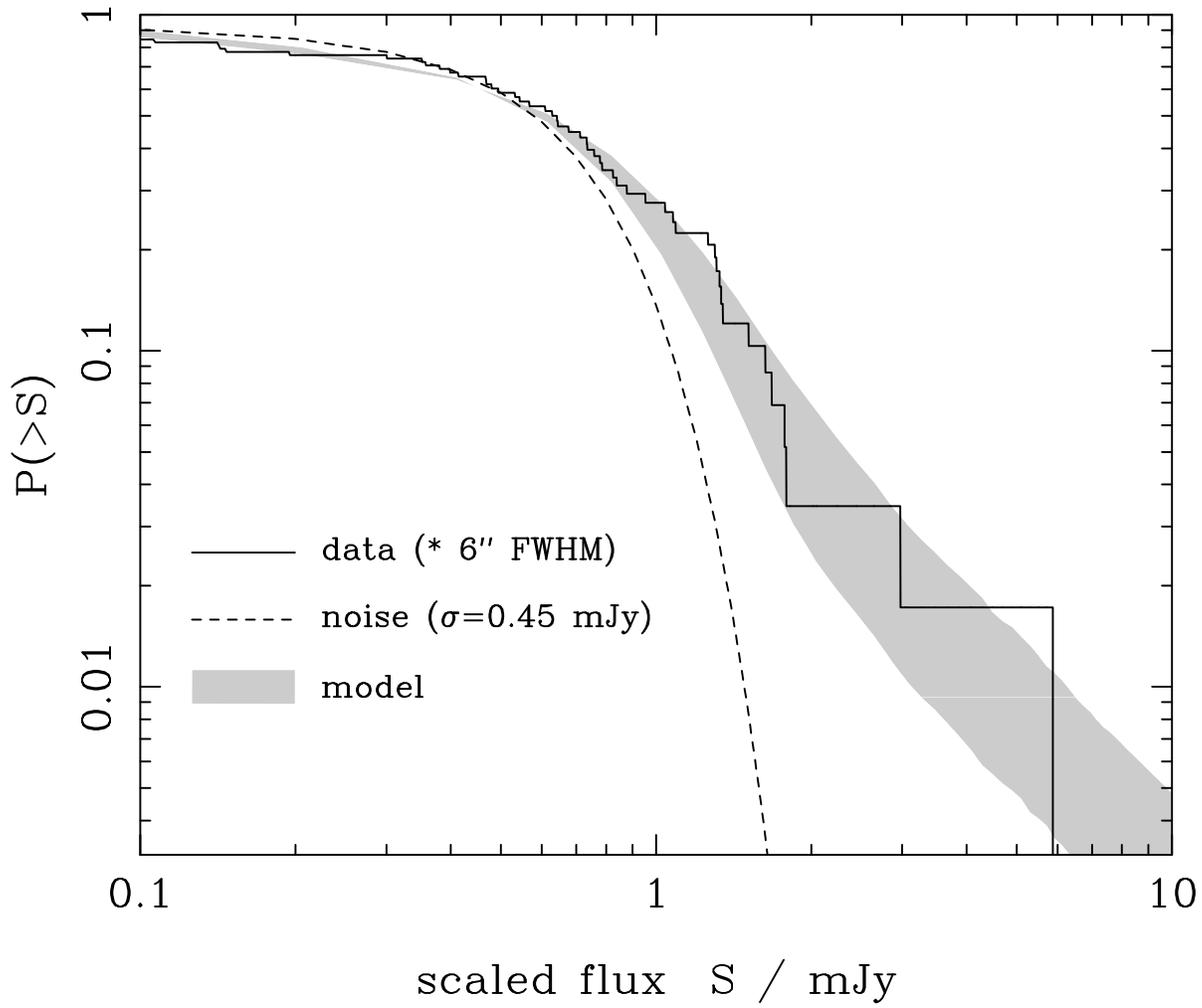
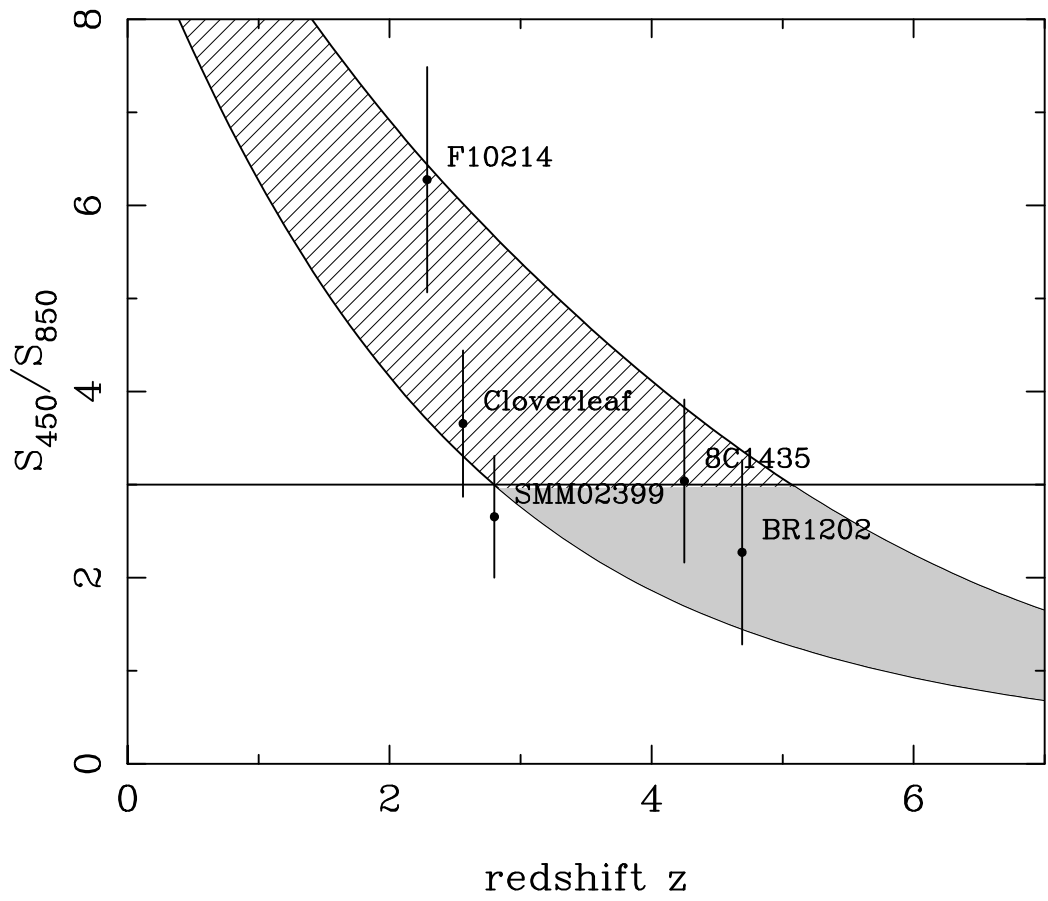
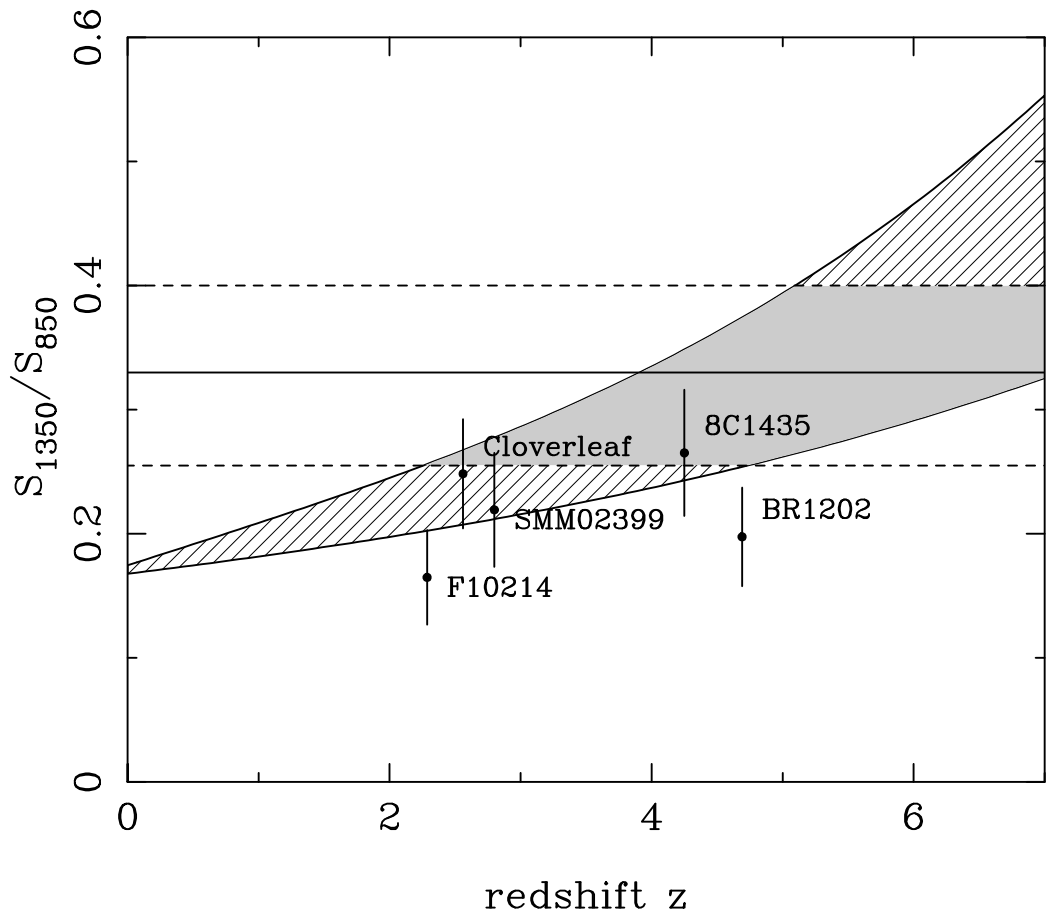
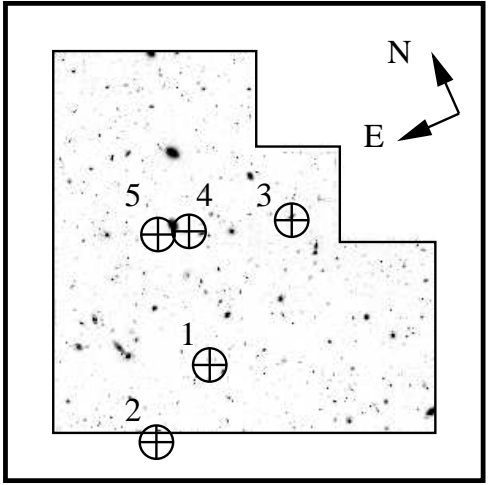
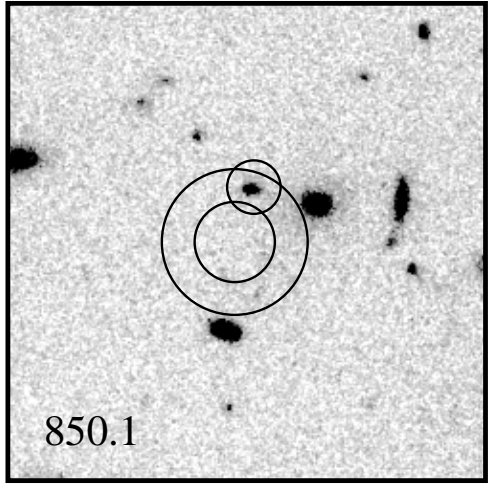
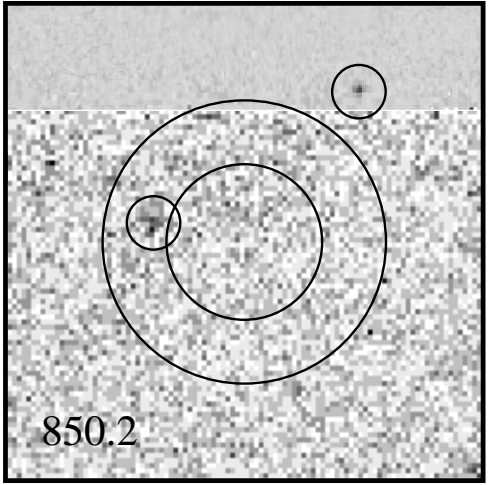
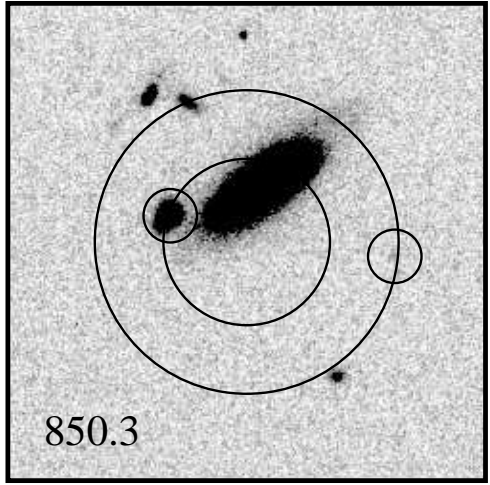
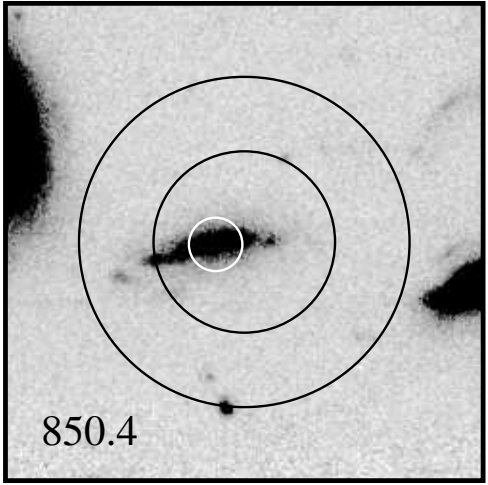
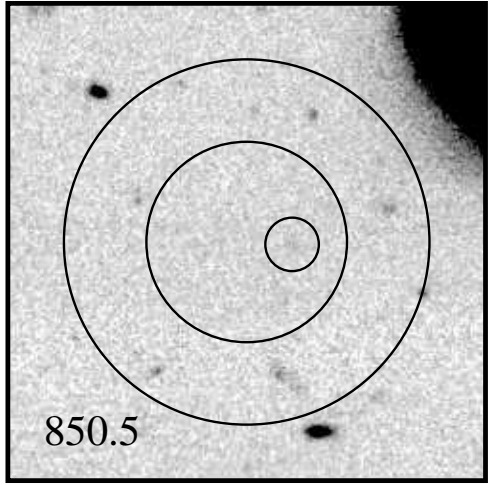
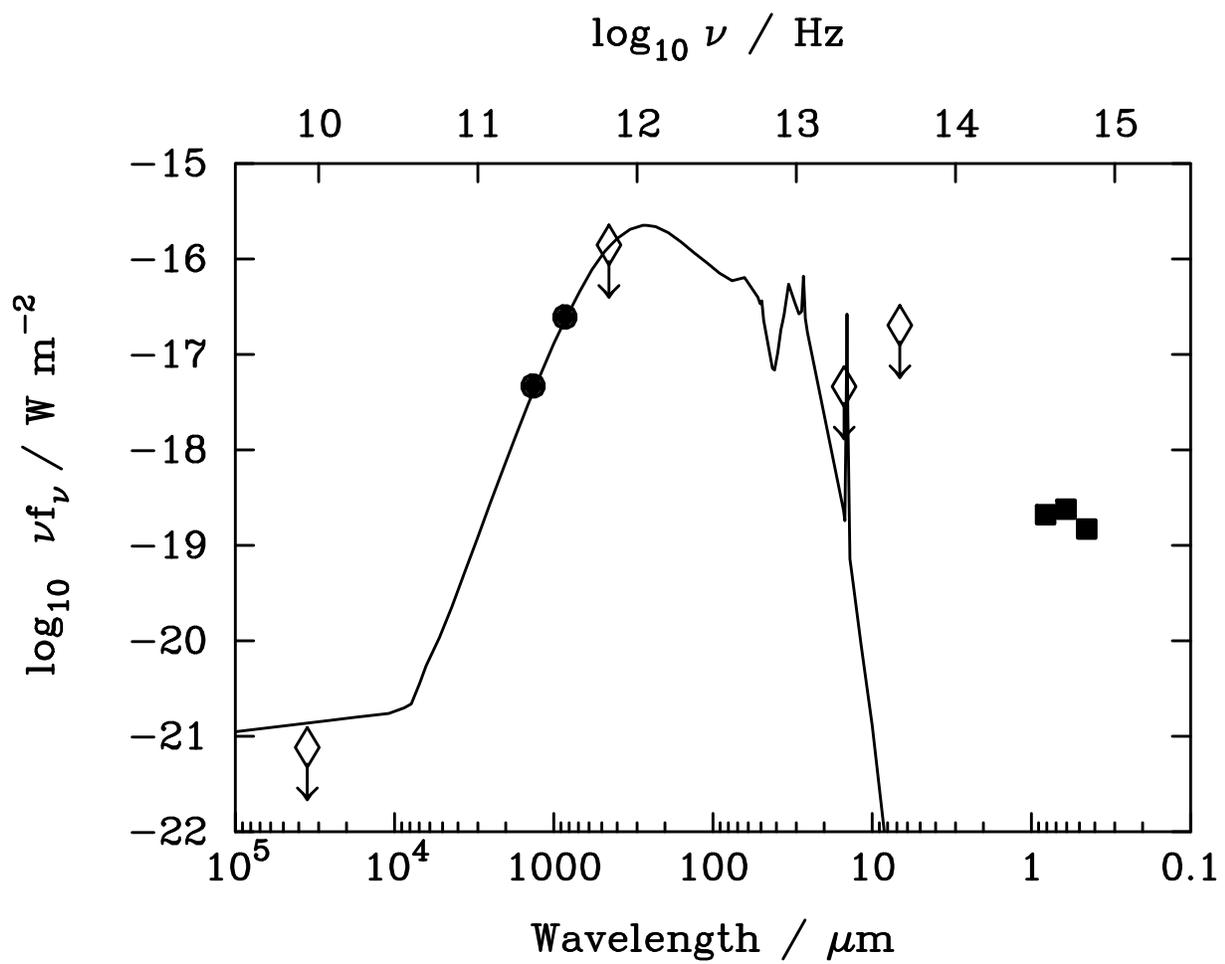
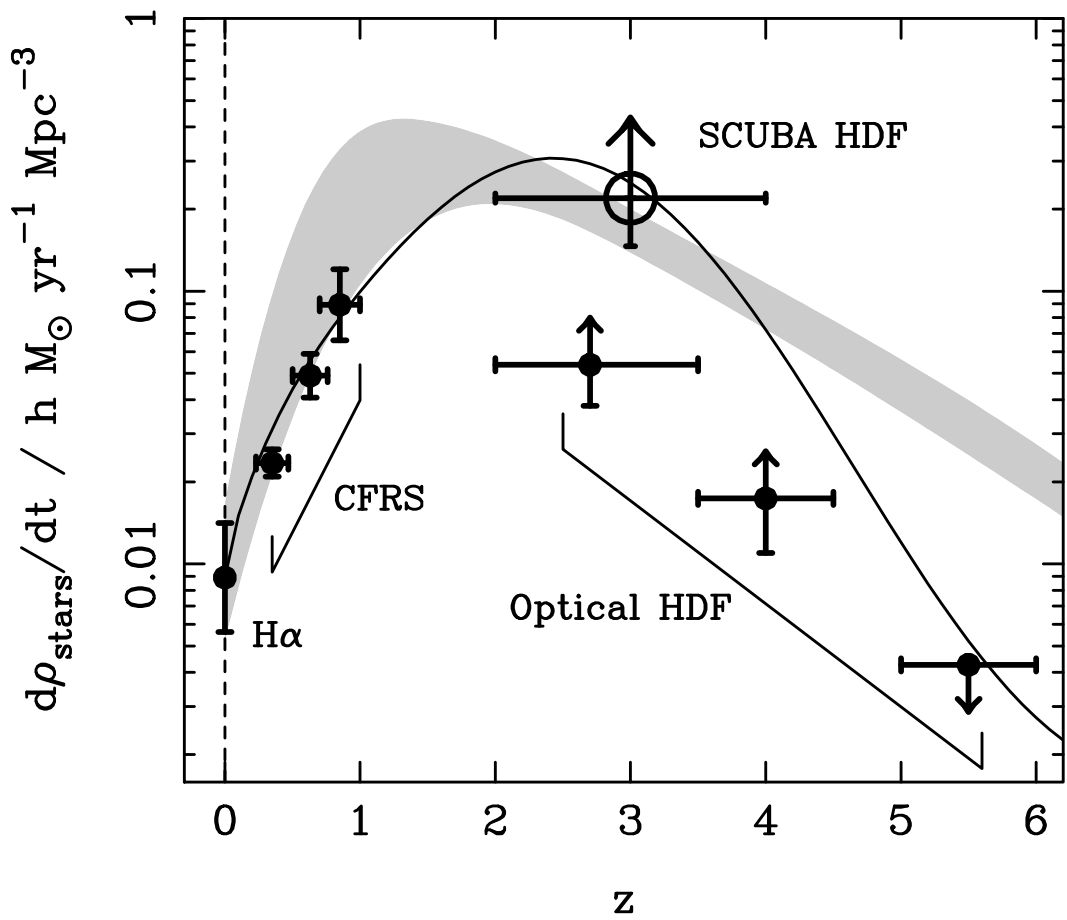
\end{document}